\documentclass[manuscript]{aastex}

\slugcomment{To appear in Astrophysical Journal Letters}

\shorttitle{Cygnus X-3's Little Friend CP}
\shortauthors{McCollough et al.}

\begin{document}



\title{Cygnus X-3: Its Little Friend's Counterpart, the Distance to Cygnus X-3, and Outflows/Jets}


\author{M. L. McCollough}
\affil{Smithsonian Astrophysical Observatory, 60 Garden Street,
    Cambridge, MA 02138, U.S.A.}
\email{mmccollough@cfa.harvard.edu}

\author{L. Corrales}
\affil{Kavli Institute for Astrophysics and Space Research, Massachusetts Institute of Technology, Cambridge, MA 02139,
U.S.A.}

\and

\author{M. M. Dunham}
\affil{Smithsonian Astrophysical Observatory, 60 Garden Street,
    Cambridge, MA 02138, U.S.A.}
\affil{Department of Physics, State University of New York at Fredonia, Fredonia, NY 14063, U.S.A.}

\begin{abstract}
Chandra observations have revealed a feature,within 16" of Cygnus X-3 which varied in phase with Cygnus X-3.  
This feature was shown to be a Bok globule which is along the line of sight to Cygnus X-3.  We report on observations 
made with Submillimeter Array (SMA) to search for molecular emission from this globule, also known as Cygnus X-3's 
``Little Friend." We have found a counterpart in both $\rm ^{12}$CO (2-1) ~and~ $^{13}$CO (2-1) emission. 
From the velocity shift of the molecular lines we are able to 
find two probable distances based on the Bayesian model of Milky Way kinematics of \cite{r3}.  For the LF velocity 
of -47.5 km/s, we get a distance of $\rm 6.1 \pm 0.6 ~kpc$ (62\% probability) and $\rm 7.8 \pm 0.6 ~kpc$ (38\% probability).  This yields a 
distance to Cyg X-3 of $\rm 7.4 \pm 1.1 ~kpc$ and $\rm 10.2 \pm 1.2 ~kpc$, respectively.  Based on the probabilities entailed, 
we take $\rm 7.4 \pm 1.1 ~kpc$ as the preferred distance to Cyg X-3.
We also report the discovery of bipolar molecular outflow, suggesting that there is active star formation occurring within the Little Friend.
\end{abstract}


\section{Introduction}

Cygnus X-3 is a high mass X-ray binary (HXRB) which lies in the Galactic plane (79.84\degr , +0.70\degr ).  Its X-ray emission is 
modulated by a factor of $\rm \sim 2.5$ on a 4.8 hour orbital period.  At an estimated distance of 
9 kpc \citep{pbpt}, it lies behind two spiral arms of the Milky Way and the Cygnus X star forming region, 
giving an opportunity to use the  X-ray emission from the HXRB to probe many types of ISM structures.  

Examination of Chandra X-ray data has shown that a small knot of X-ray surface brightness (the ``Little Friend" hereafter referred to as LF , 
see Figure 1) located 15.6\arcsec ~from the binary is modulated with the same period, but 
shifted in phase by 0.56 (\cite{msv} hereafter referred to as MSV).  The phase shift and spectrum of the feature is 
consistent with the phenomenon of X-ray scattering by  dust in the interstellar medium \citep{jo,r83}.

The angular size, spectrum, and estimated distance to the feature of 7 kpc imply that the scattering comes from a dusty $\rm 2-24 ~M_{\sun}$ 
cloud (gas + dust) confined to a region 0.2 pc across. These are the characteristics of a Bok globule \citep{br,cyh}.  Since a Bok globule is a 
small molecular cloud we might expect that it could be detected in CO emission. To see if this is possible we have observed the LF with the 
Submillimeter Array (SMA) and compare resulting observations with the {\it Chandra} X-ray observations of the LF.

\section{Observations}

\subsection{Chandra Observations}
 
 The {\it Chandra} observation used for this analysis was a 50 ksec quenched state observation 
(OBSID: \dataset [ADS/Sa.CXO#obs/6601]{6601}). During this quenched state the X-ray was high (RXTE/ASM: 2-12 keV 
count rates were $\rm \sim 25-30 ~cts~s^{-1}$) the hard X-ray very low (Swift/BAT:15-50 keV band had an average count rate of 
$\rm \sim 0.0 ~cts~s^{-1}$), and the radio low (Ryle radio telescope:15 GHz radio fluxes were $\rm \sim ~ 1~mJy$).  These values 
are all typical of a Cygnus X-3 quenched state \citep{as,mm,we}.

In order to understand the structure of the LF we used ZHTOOLS \citep{av} to create and remove the PSF of Cygnus X-3. From the right half of 
the zero-order image we created a radial profile (mkprof) and in turn used this to create a Cygnus X-3 PSF image (prof2img). We then subtracted and 
normalized the image using the PSF image (imarith). We finally mask out the readout streak to improve the dynamic range of the image.  The final 
smoothed image (3 pixel gaussian smooth) is shown in Figure 1.  One can clearly see the LF as well as an extended feature  
to the left.  This type of extension is typical of what is seen in Bok globules. A movie showing the Cygnus X-3 phase dependence of the LF has been created 
(see Figure 1). The dark vertical feature seen on the right side is instrumental (the result of a detector node boundary). 
 
In our analysis of these observations, we used version 4.3 of the CIAO tools.  The {\it Chandra}
data retrieved from the archive were processed with ASCDS version 7.7.6 or higher.

\subsection{Submillimeter Array (SMA) Observations}

Two tracks of observations of Cygnus X-3's LF were obtained with the SMA on 
2015 September 7 and 2015 September 9 in the compact configuration with seven 
antennas, providing projected baselines ranging from 6 to 74 m.  A single 
pointing was observed at a phase center of R.A.=20:32:27.1, 
decl.=$+$40:57:33.8 (J2000).  The observations were obtained with the 
230 GHz receiver tuned to approximately 230 GHz (1.3 mm).  The correlator 
was configured to provide simultaneous observations of the $^{12}$CO, $^{13}$CO, 
and C$^{18}$O J $=2-1$ lines along with the 230 GHz continuum.  
The observations were obtained in moderate weather conditions, with the 
zenith opacity at 225 GHz ranging between $0.2-0.3$ and the system 
temperatures ranging between $350-1000$ K depending on elevation.  
Regular observations of the calibration sources sources mwc349a and bllac 
were interspersed with those of the LF for gain calibration.  
3c454.3 was used for bandpass calibration, and Uranus and Neptune were used 
for absolute flux calibration.  We conservatively estimate a 20\% uncertainty 
in the absolute flux calibration.  The data were inspected, flagged, and 
calibrated following standard techniques using the MIR software 
package\footnote{Available at 
https://www.cfa.harvard.edu/$\sim$cqi/mircook.html}.  They were then imaged, 
again following standard techniques, using the 
Multichannel Image Reconstruction, Image Analysis, and 
Display (MIRIAD) software package configured for the SMA\footnote{Available at 
http://www.cfa.harvard.edu/sma/miriad/}.

Combining both tracks, the continuum observations obtained a 1$\sigma$ rms of 
1.1 mJy~beam$^{-1}$ at a central frequency of 226 GHz, with a beam size 
and position angle of 3.4$''$ $\times$ 2.5$''$ and $-$65.2 degrees, 
respectively.  The $^{12}$CO J $=2-1$ observations obtained a 
1$\sigma$ rms of 51 mJy~beam$^{-1}$ in 1.0 kms~$^{-1}$ channels, with a beam 
size and position angle of 3.3$''$ $\times$ 2.4$''$ and $-$64.4 degrees, 
respectively.  The $^{13}$CO J $=2-1$ observations obtained a 
1$\sigma$ rms of 40 mJy~beam$^{-1}$ in 2.5 kms~$^{-1}$ channels, with a beam 
size and position angle of 3.4$''$ $\times$ 2.5$''$ and $-$64.4 degrees, 
respectively.  The C$^{18}$O are not discussed here.

\section{Counterpart to Cygnus X-3's  Little Friend} 

From the SMA obsevations we found the following: 

        {\bf Confirmation of Bok globule properties:} 
	We clearly detect the LF both in $^{12}$CO (2-1) and $^{13}$CO (2-1) in the velocity range between -44 km/s and -49 km/s.	
	Since the $^{13}$CO (2-1) emission is less abundant than  $^{12}$CO (2-1), detectable emission is confined to the globule itself.  It is thus less 
	spatially extended than the $^{12}$CO (2-1) emission and less subjected to spatial filtering in the interferometer observations, and as a result it 
	is expected to be a better tracer of the globule structure.  Indeed, it closely matches the structure seen in the X-ray.
	In Figure 2 we see that the strongest $^{13}$CO (2-1) emission which is observed at -47.5 km/sec (represented by the contours) overlays the X-ray 
	emission from the LF.

	{\bf Dust cloud properties:}  
	We did not detect any continuum thermal dust emission in our observations.  
	The noise in the continuum at 1.3 mm was $\rm 1.1 ~mJy/beam$ in our observation.  
	This non-detection  of the dust continuum means that the LF must at the low mass end of what
	is expected for Bok Globules.  Using our lower flux limit we find  a $\rm 3 ~ \sigma$ mass limit 
	of $\rm 4.2 ~M_{\odot}$ (total gas+dust mass, so a dust mass upper limit of $\rm 0.042 ~M_{\odot}$ assuming a standard gas-to-dust ratio).
	This estimate is subject to factors of two uncertainty depending on the exact temperature and opacity of the dust grains.  This constrains the 
	Bok Globule to be on the lower end of the mass range inferred from the X-ray data.	
	
			
	{\bf The Distance to the Little Friend:} If we assume that the velocity shift of the CO~(2-1) line is primarily due to Galactic rotation, this will allowed us
	to pinpoint the distance to the LF. Based on a study of trigonometric parallaxes of star-forming regions, \cite{r1} provide a method and code that allows us to 
	calculate the kinematic distance to the LF.  Using the location of the LF ($\rm RA:20^{h}32^{m}27.1^{s}, DEC:+40\degr 57\arcmin 33.8\arcsec$) we calculate the 
	kinematic distance as a function of velocity using the code from \cite{r1} with updated Solar parameters values from model A5 of \cite{r2}.  We estimated the 
	error in the distance by assuming a systematic errors of $\rm \pm 7 ~km/sec$ in velocity measurements($\rm V_{LSR}$) as was done in \cite{r1}.  We find that for 
	a velocity of -47.5 km/sec the LF has a distance of ${\rm 7.62 \pm 0.62 ~kpc}$.  
	
	But \cite{xu06} have noted that parts of the Perseus arm (which is in the direction of the little friend) have shown anomalous motions which may lead to larger 
	errors in the estimated kinematic distances.  To address this we used a parallax-based distance estimator tool \footnote{Available at http://bessel.vlbi-astrometry.org/bayesian}
	 \citep{r3} which uses a Bayesian approach to account for spiral 
	arm signatures, kinematic distance estimates, Galactic latitude, and measured parallaxes to determine a source distances and its associated probabilities.  We ran 
	this tool using a velocity of -47.5 km/sec for the LF and arrive at a distance of ${\bf 6.08 \pm 0.64 ~kpc}$ with a  62 \% probability (see Figure 3).  We also find a secondary peak (38 \% probability) at
	${\bf 7.85 \pm 0.6 ~kpc}$ which corresponds to what we found from the kinematic distance estimate alone.  It should be noted that the at the largest velocity ($\rm -49 ~km/sec$) 
	at which we detect CO emission the probability for these two locations become equal.
	
	This confirming that the LF is the most distant Bok Globule known to date.  The size of central X-ray and CO~(2-1) emission of the LF is from a region of size 
	$\rm \sim 0.11 \times 0.16 ~pc$ corresponding to  $\rm (2.3 \times 3.3) \times 10^4 ~AU$.  For the distance of  6.08 ~kpc it is of interest to note that the LF will be located in the 
	middle of the Perseus arm at a location where the local branch joints it \citep{xu13}.  This is an area for which one would expect to find active star formation.

 
 \section{Distance to Cygnus X-3}
 
 Now that we have the distance to the LF we can use the X-ray small angle scattering relationship between the
 LF and Cygnus X-3 (Equation 3 of MSV)

\begin{equation}
\rm \Delta t ~=~ 1.15 \Theta^2_{obs} \frac{Dx}{1-x} ,
\end{equation}

\noindent
to determine the distance to Cygnus X-3. From MSV 
we know $\rm \Delta t = (0.56 \pm 0.02) t_{cx3}$ where $\rm t_{cx3} = 17.25~ksec$ is the observed orbital period of Cygnus X-3,
$\rm \Theta_{obs} = 15.6\arcsec$, and 
$\rm Dx = 6.08 \pm 0.64 ~(7.85 \pm 0.6) ~kpc $
is the distance to the LF.  
We find the fractional distance of the Little Friend to be 
$\rm x = 0.82 \pm 0.09 ~(0.77 \pm 0.07) ~kpc $
which in turn gives us the  
a distance to Cygnus X-3 of 
{$\rm 7.41 \pm 1.13 ~kpc ~(10.16 \pm 1.21) ~kpc $}
Based on the probabilities entailed, we take $\bf 7.4 \pm 1.1 ~kpc$ as the preferred distance to Cyg X-3.  This value is in good agreement 
with the best distance estimate of 7.2 kpc determined from the X-ray dust scattering halo \citep{lz09}.
This is one of the most precise determination of Cygnus X-3's distance to date (currently uncertainty of 7-13 kpc from \cite{pbpt}).  

With the new preferred
distance estimates to the LF and Cygnus X-3 we can reevaluate the relationship between the two (see the discussion in MSV).  Taking the error in $\Delta t$ as the main factor in determining the 
uncertainty in the relative distance between Cygnus X-3 and the LF then we find they are between $\rm 0.84 - 1.82 ~kpc$ apart.  Of the previously suggested relations the chance that this a {\it random alignment} is still a 
reasonable possibility given we are looking down the local arm and the LF is likely setting in the middle of the Perseus arm. This would also improve the {\it microquasar jet-inflated bubble} scenario as observed  by \cite{psm} 
as a possibility.  However the {\it supergiant bubble shell} would be less of a possibility since the the LF is likely in a rich star forming region and Cygnus X-3 is now well outside of the spiral arm.

But these new results lead to another interesting explanation of the relationship between Cygnus X-3 and the LF.  The progenitor of Cygnus X-3 is expected to be a binary composed of two Wolf-Rayet stars which was created in a star-forming region 
such as were the LF was found.  When an X-ray binary is formed there is strong observational evidence that it experiences a natal kick from the supernova explosion (see \cite{tw} and reference therein).  If we take the separation to be $\rm \sim 1 ~kpc$ and 
assume that the SN explosion that created Cygnus X-3 was between $\rm 1-5 \times 10^6 ~yrs$ ago (lifetime of Wolf-Rayet star) we then would get a natal kick of between $\rm 190-980 ~km/sec$ , well within the range has been observed for other X-ray binaries and 
pulsars \citep{tw}.  \cite{hsf} constrained the magnitude of systemic radial velocity of Cygnus X-3 to be $\rm < 200 ~km/sec$ making the lower value of natal kick and hence the longer time since the SN more probable.  Whereas \cite{fp} find clear indications of an 
$\rm \sim 800 ~km/sec$ redshift of Cygnus X-3's X-ray line spectra which is consistent with the upper bound of the estimated natal kick.  Thus the relationship between Cygnus X-3 and the LF 
and their separation can be explained in terms of natal kick at time of the formation the Cygnus X-3 as an X-ray binary.  This also explains why Cygnus X-3 does not appear to currently be in an area of rich star-formation.

\section{Cygnus X-3 Little Friend's Outflows/Jets:} 

A somewhat unexpected result of these observations is clear evidence of outflows/jets from the LF in both $^{12}$CO (2-1) and $^{13}$CO (2-1) (see Figures 2, 4, and 5).  Such outflows/jets are know to 
occur in molecular clouds \citep{A07, F14}.  This clearly indicates that a protostar has formed in the LF and a molecular outflow has started to occur.  

In Figure 4 we can see prominent blue and red-shifted outflows/jets coming from the LF.  The blue-shifted component (relative velocity range of 0 to -2 km/sec) extends out 14\arcsec ~($\rm \sim 0.41$ pc) with an axial ratio of $\rm \sim 2.5$.  
The red-shifted component (relative velocity of -1 km/sec) extends out 18\arcsec ~($\rm \sim 0.53$  pc) with an axial ratio of $\rm \sim 2.3$.   The blue-shifted component also shows a bend in the outflow/jet at about 7\arcsec ~($\rm \sim 0.21$  pc) from the 
LF and make a $\rm \sim 40\degr$ change in the flow. The highest observed velocity (-2 km/sec) flow appears as two knots with one located at the LF and other at the bend in the outflow/jet.    The knots may indicate the 
location of where a shock is occurring in the outflow.  From Figure 2 the $^{13}$CO (2-1) emission extends from the LF to the point at which the outflow bends.
The red-shifted component (1 km/sec) shows evidence of possible of knots near its end.  
Similar low velocity bipolar molecular outflows have been observed in other low-mass star forming clouds \citep{DM11}.
Figure 5 shows the velocity structure of the outflow/jet.  The structure that is observed may be the result of 
precession of the protostar or an interaction with the ISM. What this indicate is that we are seeing a jet + outflow (wind) from a newly emerging protostar with the possibility of shocks occurring.  

Outflows as signposts of star formation are very commonly detected.  In order to compare the outflow driven by the LF, located at a distance of 6.08 kpc, with those driven by typical low-mass 
protostars in nearby ($d < 500$ pc) Gould Belt clouds, we use the $^{12}$CO data to calculate the mass of the LF outflow.  We assume optically thin, LTE emission at an excitation temperature of 
50~K \citep{DM14}, and follow the procedures outlined in \cite{DM14} (see their Appendix C) to convert from brightness temperature to total outflow mass.  The resulting outflow mass is $\rm 0.07 ~M_{\odot}$.  
\footnote{
For the secondary distance peak the outflow would correspond to $\rm 0.11 ~M_{\odot}$}
Comparing to recent surveys of outflows driven by low-mass protostars in the nearby ($d < 500$ pc) Gould Belt clouds (e.g., \cite{CE10}; \cite{DM14}), the LF outflow is within but near the top end of the 
range of measured outflow masses ($\rm 10^{-3}$ to a few tenths of a $\rm M_{\odot}$).  Thus this detection, at 6.08 kpc, is consistent with an outflow driven by a low-mass protostar, and suggests an outflow near 
the top end of the mass scale for these types of regions.


\section{Summary}

In this study we have found the molecular counterpart to the first, and to date the most distant, Bok globule to have been seen in the X-ray.  
Also as a result of this work we have determined the most accurate distance to Cygnus X-3 to date 
($\sim$ 15\%).  Also as of result of this analysis 
we provide insight into where Cygnus X-3 likely formed and an estimate of a natal kick it received when the X-ray binary was formed via a SN explosion.
In Figure 4 we see what can best be described as a stellar cycle of life.  
We have in the X-ray from Cygnus X-3, a X-ray binary with a compact object representing an endpoint of stellar evolution.  
Also in the X-ray from the scattered X-rays of Cygnus X-3 we see a Bok globule (the LF), a small dense molecular cloud, from which stars are know
to form.  Finally in CO emission we see outflows/jets from the LF which give a clear indication that a protostar has indeed formed and is creating an outflow.

\section{Acknowledgments}

MLM wished to acknowledge support from NASA under contract NAS8-03060. This research has made use of data obtained from the 
Chandra Data Archive and software provided by the Chandra X-ray Center (CXC).  MMD acknowledges support from the Submillimeter Array (SMA) through an SMA 
postdoctoral fellowship, and from NASA through grant NNX13AE54G.  Support for the work by LRC came in part from the NASA Earth and Space Science Fellowship Program,
grant NNX11AO09H.  This work is based primarily on observations  obtained with the Submillimeter Array, 
a joint project between the Smithsonian Astrophysical Observatory and the Academia Sinica Institute of Astronomy and Astrophysics and funded by the Smithsonian 
Institution and the Academia Sinica. Special thanks to J.DePasquale for the help with the composite X-ray and CO images.  
We wish to thank the referee for pointing 
out several important papers which have improved this work.

\begin{figure}[ht!]
\begin{center}
\includegraphics[width=1.0\columnwidth]{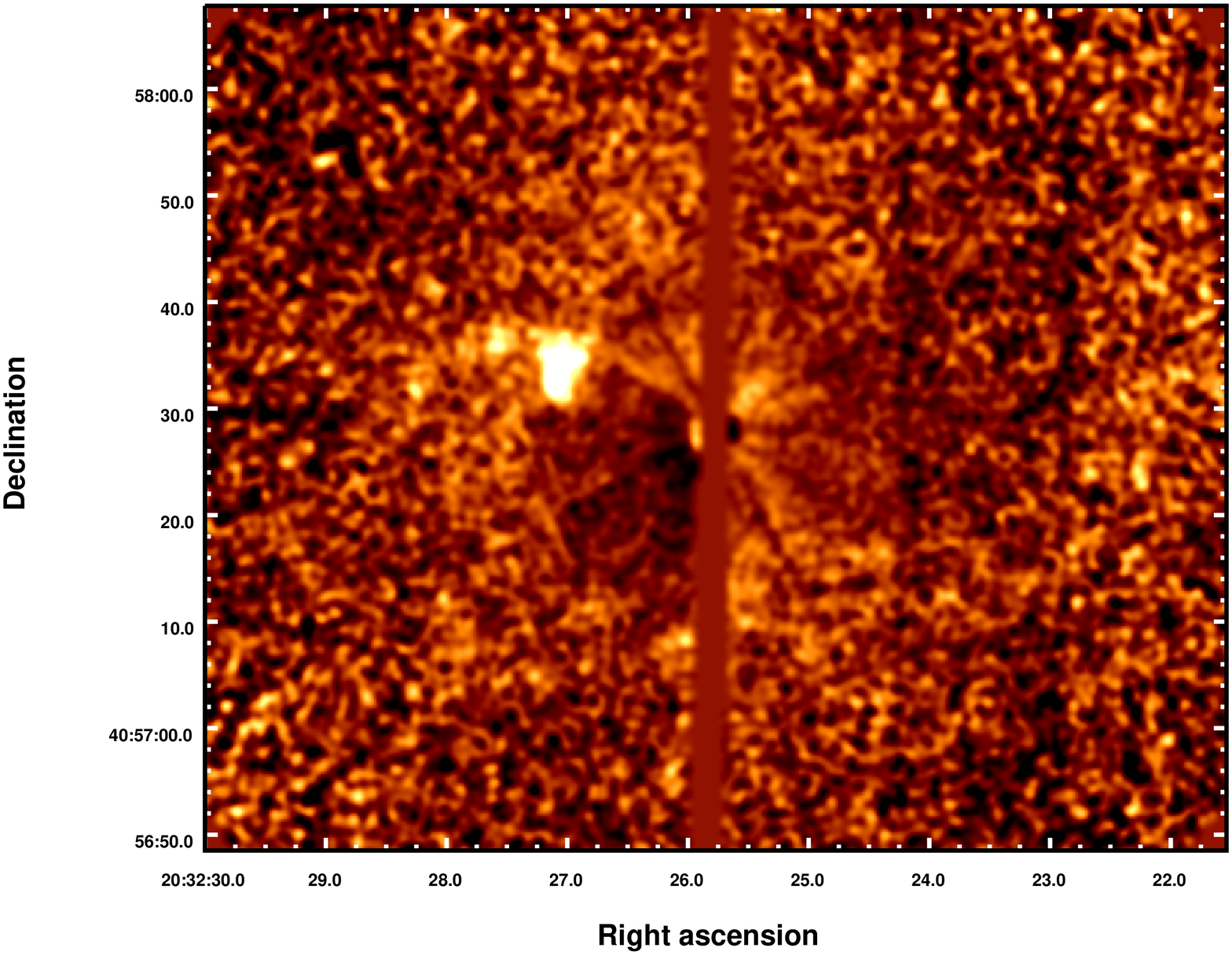}
\caption{This is a smoothed X-ray (1-8 keV) image of the Little Friend with Cygnus X-3's PSF removed and the readout streak mask out.  One can clearly see the LF and that it appears to 
lie along an arc with two knots of emission further out.
{\bf Online electronic edition:} To better visualize the phase relationship between the feature and it structure relative to Cygnus X-3 a movie 
was created. The method used is the same as was described in MSV except the Cygnus X-3 PSF subtraction technique described in section 2.1 was used for each frame of the movie. The movie goes 
through a full Cygnus X-3 orbital period starting at phase 0.0.
}
\end{center}
\end{figure}

\begin{figure}[ht!]
\begin{center}
\includegraphics[width=1.0\columnwidth]{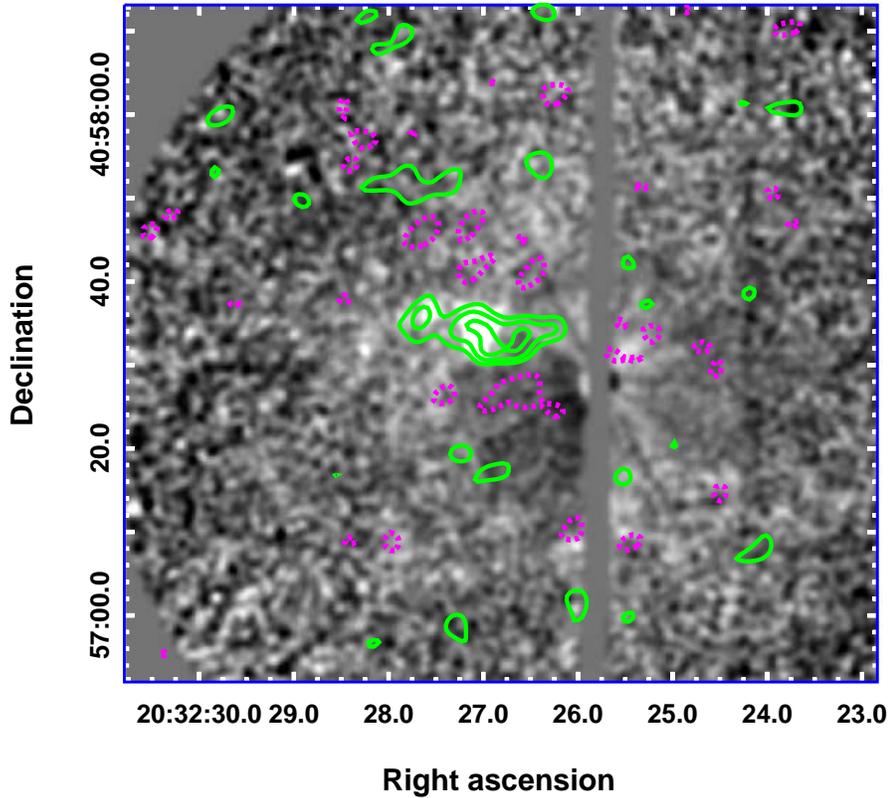}
\caption{This is X-ray emission seen in Figure 1, for the SMA field, with contours ([-2, 2, 3, 4] $\times$ 40 mJy/beam) of $\rm ^{13}$CO (2-1) at -47.5 km/sec overlayed (green for positive values and dotted magenta for negative).
 This shows a clear association of CO emission with the LF.  The CO clearly matches the X-ray emission from the LF including the first blob to the left.  Also note the extension of the  $\rm ^{13}$CO (2-1) to the right away from the LF.
}
\end{center}
\end{figure}

\begin{figure}[ht!]
\begin{center}
\includegraphics[width=1.0\columnwidth]{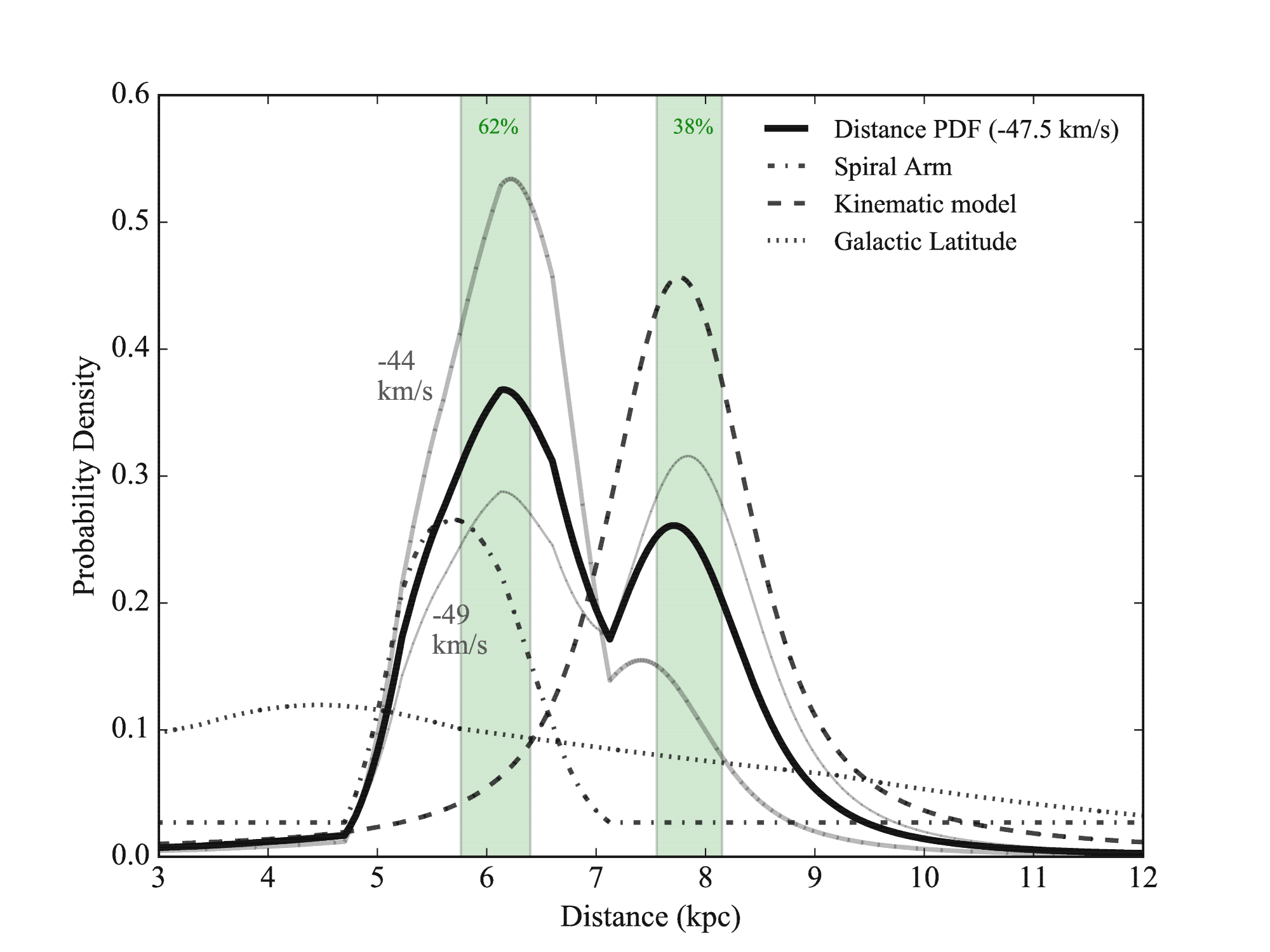}
\caption{This is a plot of probability density as a function of distance for Cygnus X-3 using a $\rm V_{LSR}$ of $\rm -47.5 ~km/sec$.  The values come from the Bayesian distance estimator of \cite{r3}.  The thick black solid line is the joint 
probability density. The dash-dot, dashed,and dotted lines are the principle components in the model (spiral arm, kinematic distance, and Galactic latitude respectively).  The green shaded areas represent possible locations (with probabilities) and whose 
width represents $\rm 1 \pm ~\sigma$ uncertainties in location.  The thick and thin gray lines represent the probability densities for the minimum and maximum $\rm V_{LSR}$ observed for CO emission.
%
}
\end{center}
\end{figure}

\begin{figure}[ht!]
\begin{center}
\includegraphics[width=0.9\columnwidth]{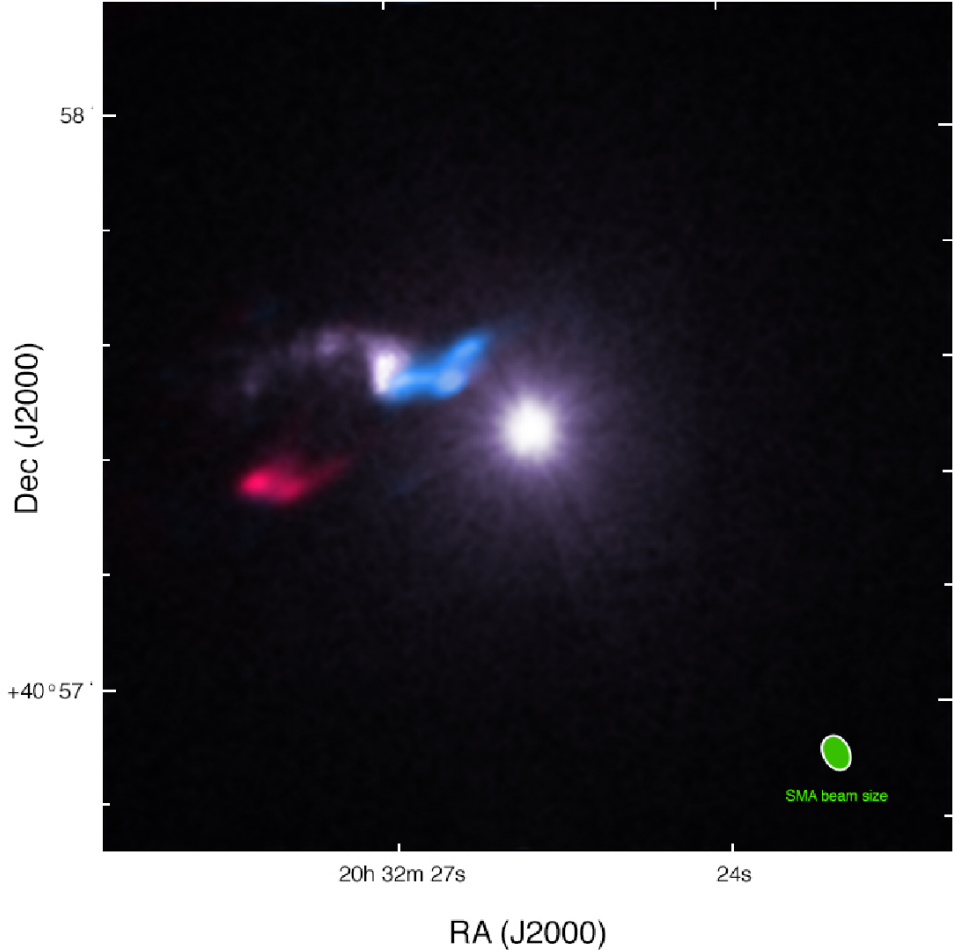}
\caption{This is a composite image created from the X-ray (1-8 keV) Chandra data (purple/white) and CO emission obtained from the SMA showing the outflow/jet from the LF.  
In the X-ray you can see Cygnus X-3 and the LF.  The blue is $\rm ^{12}$CO (2-1) emission with negative velocities (0 to -2 km/sec) relative the strongest $\rm ^{13}$CO (2-1) 
which we believe represents the rest velocity of the globule.  The red is $\rm ^{12}$CO (2-1) emission from positive velocities (1 km/sec) we believe are associated with the outflow/jet.
}
\end{center}
\end{figure}

\begin{figure}[ht!]
\begin{center}
\includegraphics[width=0.9\columnwidth]{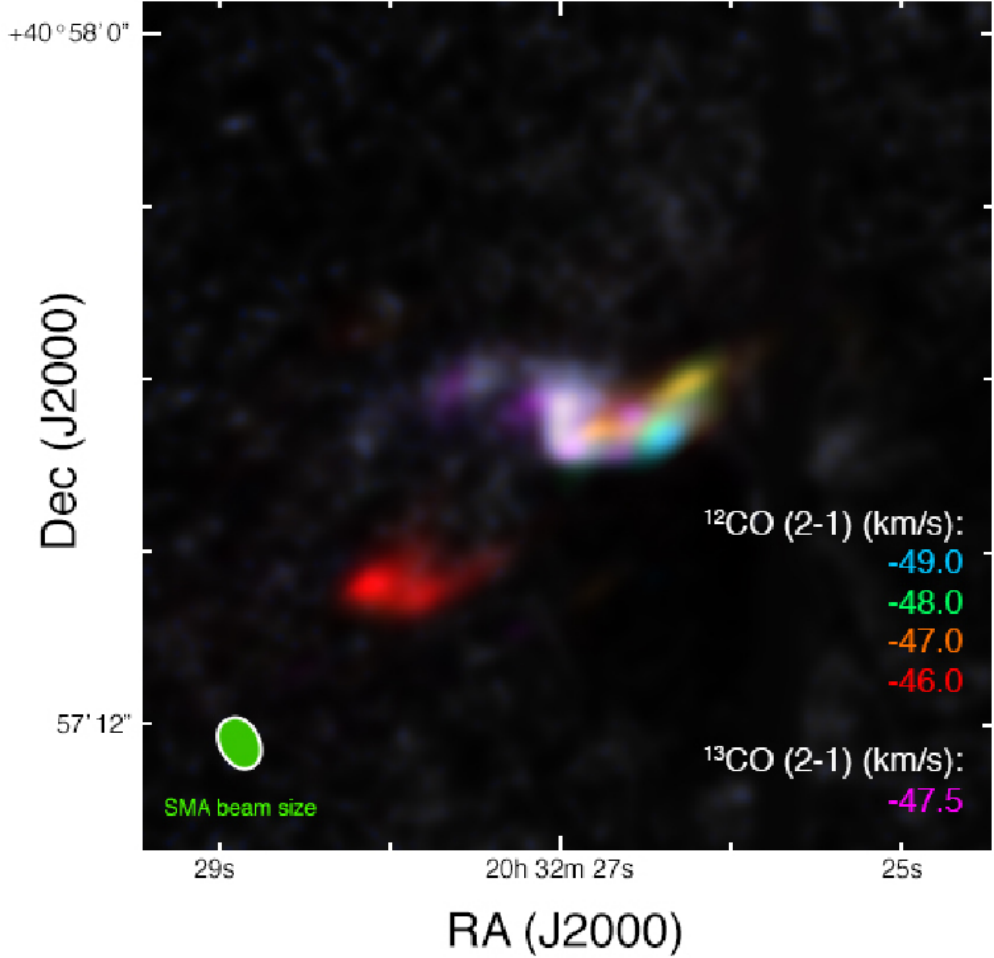}
\caption{This Cygnus X-3 PSF subtracted image (purple/white) of the LF and the CO emission.  The magenta is $\rm ^{13}$CO (2-1) which covers the LF as well as part of the jet.  
The other colors represents the various velocity components of emission from $\rm ^{12}$CO (2-1) showing the velocity structure of the jets which is discussed in the text.
}
\end{center}
\end{figure}

\end{document}